# Recent trends and analysis of Generative Adversarial Networks in Cervical Cancer Imaging

Tamanna Sood[1]
Department of Computer Science and Engineering
Punjab Engineering College (Deemed to be University)
Chandigarh, India
tamanna.sood1893@gmail.com

Dr. Rajesh Bhatia[2]
Department of Computer Science and Engineering
Punjab Engineering College (Deemed to be University)
Chandigarh, India

Dr. Padmavati[2]
Department of Computer Science and Engineering
Punjab Engineering College (Deemed to be University)
Chandigarh, India

*Abstract*— **Cervical cancer is one of the most common types of cancer found in females. It contributes to 6-29% of all cancers in women. It is caused by the Human Papilloma Virus (HPV). The 5-year survival chances of cervical cancer range from 17%-92% depending upon the stage at which it is detected. Early detection of this disease helps in better treatment and survival rate of the patient. Many deep learning algorithms are being used for the detection of cervical cancer these days. A special category of deep learning techniques known as Generative Adversarial Networks (GANs) are catching up with speed in the screening, detection, and classification of cervical cancer. In this work, we present a detailed analysis of the recent trends relating to the use of various GAN models, their applications, and the evaluation metrics used for their performance evaluation in the field of cervical cancer imaging.**

## I. INTRODUCTION

Cervical cancer is ranked second among women's cancers. ~96.9k new cervical cancer cases and ~60k deaths were reported in 2020 [1] in India. Cervical cancer is caused by the Human Papillomavirus (HPV). The pap-smear test is commonly used to screen for cervical cancer. It looks for and identifies any abnormal or precancerous cells present in the cervix area[2]. If one tests positive in the pap-smear screening test, they must undergo either colposcopy or biopsy to ascertain the presence of cancer. Just like any other cancer, early diagnosis and detection of cervical cancer greatly improve the chances of successful treatment of cancer and the survival rate of the patient.

## II. LITERATURE REVIEW

Over the past few years, with the advancement in technology and artificial intelligence, many researchers have worked on the detection of cervical cancer with the help of deep learning techniques and algorithms. [3], [4], [5], [6], [7], [8] have proposed and implemented state-of-the-art models and techniques for the detection, diagnosis, and classification of cervical cancer using various deep learning algorithms. [9] and [10] have provided a detailed analysis of the deep learning algorithms that are being used for the diagnosis of cervical cancer. Apart from these algorithms, the use of Generative Adversarial Networks is also growing in the field of cancer detection. This work provides a complete analysis of the various GAN models, their uses, and evaluation metrics that have been used in the field of cervical cancer detection. [11] provides a holistic view of the various applications of GANs relating to cancer imaging. To the best of our knowledge, no such work focusing on cervical cancer exists as of now. The following sections will explain the various GAN models that have been used, the purpose of using the model, and the parameters on which the performance of the models has been evaluated.

## III. METHODS

Generative Adversarial Networks or GANs are a deep learning approach toward generative modeling. GANs use an unsupervised approach to generate an output that is unknown and based on the input [12]. They provide a solution to domain-specific data augmentation and problems that require a generative solution such as image-to-image translation.

### A. Basic architecture of GANs

The basic architecture of GANs comprises a generator module (G) which is a convolutional neural network and a discriminator module (D) which is a deconvolutional neural network. The goal of the generator model is to fool the discriminator by creating images that look like the real images while the goal of the discriminator is to identify the fake images from the real ones [13]. Figure 1 displays the architecture of a basic GAN model.

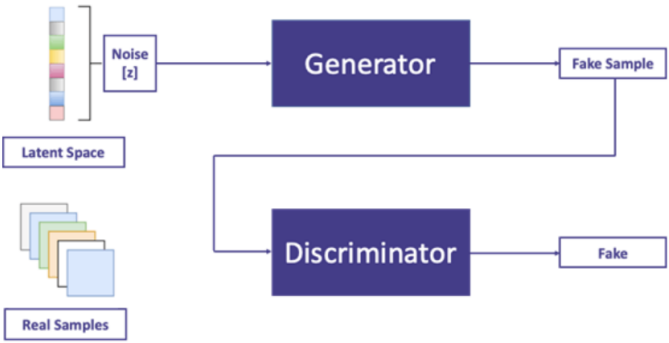

Figure 1 Architecture of a basic GAN model

Many variations of GAN have been used in the works studied. Faster Small Object Detection Neural Network (FSOD-GAN) which is a combination of the Faster RCNN and GAN has been used by [14], Residual Network and generative adversarial network (RCGAN) has been used by [15], Synergetic multiplex network (SMN) based on cycle-GAN has been proposed by [16], [17] have implemented CycleGAN, pix2pix, ESRGAN, and pix2pixHD GAN networks, MC-deblurGAN has been proposed by [18], cell-GAN has been proposed by [19] while [20] have used conditional GANs. These architectures have been discussed in the following section.

### B. Different GANs used

As discussed in the above section, many different architectures have been used by different authors in their work. This section tries to explain the various architectures that have been mentioned in the above section. The most commonly used architectures are compared in Table 1.

Table 1 Comparison of the most commonly used GAN architectures

| GAN Architecture | Paper | Type of Model | Input | Output | Loss Function | Use |
|---|---|---|---|---|---|---|
| Faster Small Object Detection Neural Network (FSOD-GAN) | [14] | Hybrid | Colposcopy Images | Class label, annotation of cervical spotting, and segmented image | $\mathcal{L}_D = -\log D(c_2) - \log(1 - D(c_2 + G(c_2\|f)))$  $\mathcal{L}_G = -\log(D(c_2 + G(c_2\|f)))$ | Segmentation and classification |
| Residual Network and generative adversarial | [15] | Hybrid | Random Gaussian noise | Synthetic dataset | Least Square Loss and Gradient Loss | Data Augmentation |
| network (RCGAN) | | | | | | |
| Synergetic multiplex network (SMN) | [16] | Hybrid | MRI Images | Segmented images of different views | Cycle consistency loss | Multi-organ segmentation |
| Cycle GAN, Pix2Pix GAN, Pix2PixHD GAN, ESRGAN | [17] | Basic | Light microscopy images | High-resolution images | Combination of perception loss, SSIM loss, content loss, and adversarial loss | Super-resolution and classification |
| MC-deblurGAN | [18] | Hybrid | Blur mobile phone cervical images | Deblurred images | Combination of adversarial loss and content loss | Image deblurring (Super-resolution) |
| Cell-GAN | [19] | Hybrid | Cell image | Cropped/ Segmented Image | $L_2$ loss function | Segmentation |
| Conditional GAN | [20] | Hybrid | Noise vector and label condition vector | Synthetic dataset | Wasserstein Gradient Penalty (WGAN-GP) Loss | Data Augmentation |

Here, the type of model states whether a state-of-the-art model has been used as is (Basic), or it has been modified or combined with some other models (Hybrid). The Input column represents the initial input to the GAN model (input to the generator module), while the Output column represents the final output of the GAN model (output of the Discriminator model). The Loss Function represents the type of loss function that has been used in the GAN model. The use column states the use for which the GAN model has been implemented. The next section further illustrates and explains the use and evaluation metrics for GANs.

## IV. RESULTS

### A. Use of GANs

As stated by [11], GANs provide a wide range of applicability when it comes to working with cancer images. The few applications for which GANs have been used in cervical cancer imaging have been explained as follows:

1. **Classification:** GANs make use of the synthetic data that has been generated for better image classification. In [14], FSOD-GAN has been used to classify the subtype of cancer with the help of digital colposcopy images. [17] classifies microscopic cell images of HPV viruses with the help of CycleGAN, Pix2pix, ESRGAN, and Pix2pixHD.
2. **Image Segmentation:** Image segmentation helps us extract our required region of interest in a particular image. GANs help in automated and unsupervised image segmentation. FSOD-GAN[14], SMN[16], and Cell-GAN[19] have been used for the segmentation of cervical cell images.
3. **Data Augmentation:** Data augmentation is the process of increasing the dataset by fabricating synthetic data that is as close to the original data in nature. Data augmentation is the original forte of GANs. RCGAN[15] and cGAN[20] have been used for the expansion of the dataset.

4. **Super-resolution:** The objective behind super-resolution is to recover finer texture from the images when it is upscaled so that the quality of the image is not compromised. Super-resolution can also help in the deblurring of images. CycleGAN, Pix2pix, ESRGAN, and Pix2pixHD have been used for the super-resolution of microscopic images[17]. MC-deblurGAN has been sued for the deblurring of mobile phone cervical images[18].

### B. Evaluation metrics for qualitative and quantitative analysis

Once the task is performed, the next step is to evaluate the performance of the model. Since there is no objective loss function in GANs, we cannot objectively assess the performance of the model [21]. Thus, several qualitative and quantitative techniques have been designed to assess the quality and diversity of the generated images. These techniques have been discussed below:

1. **DICE score:** The Dice score or the Dice coefficient (aka F1 score) corresponds to a perfect pixel match between the output and the ground truth annotation. It is used as an evaluation metric in segmentation tasks. The value of the dice score ranges from 0 to 1.
2. **Fréchet Inception Distance (FID):** FID is to evaluate the quality of generated images [22]. It is a method for comparing the statistics of two distributions by computing the distance between them. Instead of comparing the images pixel by pixel, FID compares the mean and standard deviation, and mean of the InceptionV3 layer of the CNN. The lower the FID, the better the performance of the GAN.
3. **Peak Signal to Noise Ratio (PSNR):** PSNR is an image quality metric. It is measured as the ratio between the maximum signal and the maximum noise. It is used as a quality measurement between the original image and the generated image. The PSNR value is directly proportional to the quality of the generated image [23].
4. **Perception-based Image Quality Evaluator (PIQE):** PIQE calculates the no-reference image quality score for an image using a perception base image quality evaluator [24]. The PIQE value is inversely proportional to the perceptual quality of the image.
5. **Structural Similarity Index (SSIM):** SSIM is used to calculate the similarity of two images. It measures image quality degradation [25]. It is based on visible structures in the image. It does not judge which of the two images is better.

Table 2 lists the various studies that have been discussed in this work along with all the parameters that have been discussed. It provides a complete and comprehensive comparison of the various types of GANs and how they have been used in cervical cancer analysis. Not much work has been done in the field of cervical cancer imaging using GANs, thus, only 8 studies were compared. Table 1 gives a comparative analysis of the various GAN models and their uses in cervical cancer imaging.

Table 2 Comparison of the uses and evaluation metrics used for various GAN models

| Paper | Type of dataset | Use of GAN | Model used | Evaluation metrics | Result |
|---|---|---|---|---|---|
| [14] | Colposcopy Images | Segmentation and Classification | FSOD-GAN | Accuracy | Accuracy: 98.55% |
| [15] | Cervical single-cell images | Data Augmentation | RCGAN | Fréchet Inception Distance (FID | FID score: 13.004 |
| [16] | Multi-view cervical MRI | Image segmentation | SMN | DICE Score, PSNR, MSE | DICE:0.9575, MSE:0.0024, PSNR:26.1782 |
| [17] | Light microscopy images | Super-resolution and classification | CycleGAN, Pix2pix, ESRGAN, Pix2pixHD | Specificity and sensitivity | CycleGAN: 78.81% Pix2pix: 76.37% ESRGAN: 76.62% Pix2pixHD: 84.71% |
| [18] | Mobile Phone cervical images | Image deblurring (Super-resolution) | MC-deblurGAN | PSNR, PIQE, SSIM | Avg PSNR: 36.9 Avg SSIM: 0.92 Avg PIQE: 13.1 |
| [19] | Single-cell images | Image segmentation | Cell-GAN | DICE Score | DICE: 0.943 |
| [20] | Cervical histopathology images | Data Augmentation | Conditional GAN | Accuracy | Accuracy: 71.7% |

## V. DISCUSSION

This study discussed the use of GANs in cervical cancer detection and classification. Also, the different models have been investigated and summarized that are explored in the existing literature. It can be deduced that GAN-based solutions to the classification and detection of cervical cancer are following the same trend as that of analysis of other cancer imaging using GANs [26]. GANs have been used in segmentation, classification, image augmentation, and super-resolution of images. As can be inferred from Table 1, the performance is also up to the mark. We further plan to extend our work in the same area and work on various uses of GANs for cervical cancer imaging.

## VI. CONCLUSION

We can safely say that GANs are one of the most versatile models that are currently being used. They can be used for classification, segmentation, augmentation, super-resolution, and other activities. The main advantage of GANs is the data privacy it offers. The main disadvantage however is that they are susceptible to adversarial attacks. This problem is very critical when it comes to working with medical images. Many researchers are working on finding the solution to this problem. Apart from these issues, GANs provide an open playfield in the area of image analysis. Their use has been of great advantage in the medical imaging and analysis domain. We plan to work in the same field, extending our work on GANs in the field of cervical cancer imaging.